\documentclass[%
 aip,
 amsmath,amssymb,
 reprint,%
]{revtex4-1}

\usepackage{graphicx}
\usepackage{dcolumn}
\usepackage{bm}

\usepackage[utf8]{inputenc}
\usepackage[T1]{fontenc}
\usepackage{mathptmx}
\usepackage{etoolbox}
\usepackage{color}

\makeatletter
\def\@email#1#2{%
 \endgroup
 \patchcmd{\titleblock@produce}
  {\frontmatter@RRAPformat}
  {\frontmatter@RRAPformat{\produce@RRAP{*#1\href{mailto:#2}{#2}}}\frontmatter@RRAPformat}
  {}{}
}%
\makeatother
\begin{document}

\preprint{AIP/123-QED}

\title{Bubble lifetimes in DNA gene promoters and their mutations affecting transcription}
\author{M.~Hillebrand}
\affiliation{Nonlinear Dynamics and Chaos Group, Department of Mathematics and
Applied Mathematics, University of Cape Town, Rondebosch 7701,
South Africa}
\author{G.~Kalosakas}%
 \email{georgek@upatras.gr}
\affiliation{Department of Materials Science, University of Patras, GR-26504 Rio, Greece}

\author{A.R.~Bishop} \affiliation{Los Alamos National Laboratory,
Los Alamos, NM, 87545, United States}

\author{Ch.~Skokos}
\affiliation{Nonlinear Dynamics and Chaos Group, Department of Mathematics and
Applied Mathematics, University of Cape Town, Rondebosch 7701,
South Africa}

\date{\today}

\begin{abstract}
Relative lifetimes of inherent double stranded DNA openings with lengths up to ten base pairs are presented for different
gene promoters and corresponding mutants that either increase or decrease transcriptional activity, in the framework
of the Peyrard-Bishop-Dauxois model. Extensive microcanonical simulations are used, with energies corresponding
to physiological temperature.
The bubble lifetime profiles along the DNA sequences demonstrate a significant reduction of the average lifetime
at the mutation sites when the mutated promoter decreases transcription, while a corresponding enhancement of the
bubble lifetime is observed in the case of mutations leading to increased transcription.
The relative difference of bubble lifetimes between the mutated and the wild type promoters at the position of mutation
varies from 20\% to more than 30\% as the bubble length is decreasing.
\end{abstract}

\maketitle

\section{Introduction} 
\label{sec:introduction}
As more and more research is carried out on DNA functional activity, there is increased evidence that local base pair fluctuational openings may be one important factor in the process of transcription \cite{sobel,Choi2004,Alexandrov2010}.
Hydrogen bond vibrations and local disruptions, leading to transient separations of complementary strands, have been experimentally
studied in different time scales  \cite{imino87,Krichevsky2003,HippelMarcus2013,Jimenez2016}.
Furthermore, various theoretical models have addressed dynamical and statistical properties of base pair stretchings in the double helix
\cite{Yakushevich,Peyrard1989,PBD,DPB93,DP95,Barbi99,Cocco00,Metzler03,bubble07,falo10,Manghi2016,zoli18,zoli21}.

An apparent correlation between thermally induced large local openings (``bubbles'') and functional sites along
the DNA sequence that are relevant for transcription initiation has been noted in a number of investigations \cite{Choi2004,Kalosakas2004,Alexandrov2009,Alexandrov2010,Apostolaki2011,faloPRE12,huangJBE,Nowak13,faloPLOS},
where it has been found that bubbles form with a greater probability at transcriptionally active sites than elsewhere along the promoter sequence.
These works have considered thermal equilibrium or out-of-equilibrium dynamical properties of various gene promoter segments,
using the Peyrard-Bishop-Dauxois (PBD) coarse-grained model \cite{PBD} of DNA or extensions of this model.

Examining the dynamics of DNA promoters through Langevin molecular dynamics simulations, the lifetimes of these bubble openings have been probed as potential indicators for the initiation of transcription \cite{Alexandrov2006,Alexandrov2009,Alexandrov2010}.
In particular, the adeno-associated viral (AAV) P5 promoter was shown to exhibit relatively long-lived bubbles with characteristic lengths near the transcription start site (TSS) \cite{Alexandrov2006}.
A broader study of an array of mammalian gene promoters showed the same dynamical signatures; bubbles form with greater frequency and longer lifetimes at active sites in the promoter and particularly at the TSS \cite{Alexandrov2009}.
The importance of the lifetime as an indicator of transcriptional activity was underlined by the findings that long-lived bubbles formed in functional regions, even when the occurrence probability at those regions was not exceptional.
This motif of dynamics-driven transcription was further established by the detailed analysis of the SCP1 so-called ``superpromoter'', which revealed that not only are long-lived bubbles at the TSS correlated with transcriptional activity, but mutations which significantly reduce the transcription also decrease the occurrence and the lifetimes of bubbles at the TSS \cite{Alexandrov2010}.
Consequently there is a strong interest in further investigating this relationship between bubble lifetimes and transcription, using extensive numerical simulations under various conditions of the examined system in order to deepen our understanding of dynamical
features with a possible biological role.

In this work we present a detailed analysis of {\it inherent} bubble lifetimes (i.e.~without any environmental factors or effects
from the surroundings, depending only on constant-energy fluctuations in the framework of a microcanonical evolution) across the sequence of different promoters -- a viral, a bacterial, and a very strong artificial promoter -- as well as comparisons with a corresponding mutant in each case that is known to either reduce or enhance transcriptional activity.
Following a recent study of bubble lifetime distributions \cite{Hillebrand2020},  we use the PBD model \cite{PBD} of DNA
with sequence-dependent stacking parameters \cite{Alexandrov2009ePBD}, to perform constant-energy simulations
(complementing earlier works implementing Langevin dynamics \cite{Alexandrov2006,Alexandrov2009,Alexandrov2010}),
in order to more closely probe the internal characteristic times of double strand openings without having to impose
artificial time scales through arbitrary friction coefficients.

We examine three promoters, the viral AAV P5 promoter, the bacterial Lac operon promoter, and the artificial SCP1 superpromoter
(see Section~\ref{sec:modelling_and_numerical_methods} for their sequences),
as well as one particular mutant of each case exhibiting altered transcriptional activity.
The P5 promoter is critical to the genetic activity of AAV DNA, by directing relevant expression patterns \cite{Tratschin1984}.
We also consider a double mutation of this promoter, resulting in loss of transcription activity \cite{Choi2004}.
The Lac operon promoter is a thoroughly studied regulatory region in the \textit{E. coli} K-12 bacterium. We investigate the wild type
as well as the mutant Lac UV5, which exhibits increased transcription with no need of activator action \cite{Gilbert1976}.
The final promoter is the artificially constructed SCP1 superpromoter, designed to have exceptional transcriptive behavior \cite{Juven2006}. The mutant studied here is the m1SCP1 sequence, resulting in reduced transcription \cite{Alexandrov2009}. 

The next Section~\ref{sec:modelling_and_numerical_methods} introduces the PBD model, presents the studied promoter
sequences, and outlines the methods used in the analysis of the simulation data.
Section~\ref{sec:promoter_lifetimes} lays out the results of our investigation, including a discussion of the relative lifetime changes
in the mutated promoters.
Finally, Section~\ref{sec:conclusions} summarises and concludes our work.

\section{Modelling and Numerical Methods} 
\label{sec:modelling_and_numerical_methods}
In the PBD framework considered here, a coarse-grained model is used for the base pair stretchings
and the force fields of the system are approximated through appropriate analytical potentials.
The PBD model provides a Hamiltonian for the dynamics, with the on-site intra-base-pair interaction
accounted for by a Morse potential $V$,
\begin{equation}
    \label{eq:Morse}
    V(y_n) = D_n\left(e^{-a_n y_n} - 1\right)^2,
\end{equation}
where $y_n$ represents the relative displacement from equilibrium of the bases within the $n^{th}$ base pair of a DNA sequence.
The site-dependent parameters $D_n$ and $a_n$ distinguish adenine-thymine (A-T) and guanine-cytosine (G-C) base pairs
along the sequence. 

An anharmonic coupling $W$ models the stacking energy,
\begin{equation}
    \label{eq:PBDStack}
    W(y_{n},y_{n-1}) = \frac{K_{n,n-1}}{2}\left(1+\rho e ^{b(y_n + y_{n-1})}\right)\left(y_n-y_{n-1}\right)^2.
\end{equation}

The total Hamiltonian of a DNA sequence having $N$ base pairs reads
\begin{equation}
    \label{eq:PBDHamiltonian}
    H = \sum_{n=1}^N  \left[ \frac{p_n^2}{2m}+V(y_n) + W(y_n,y_{n-1}) \right],
\end{equation}
where  $p_n$ are the conjugate momenta to the canonical displacements $y_n$. 
Periodic boundary conditions have been considered here.

Apart from the sequence-dependent spring constants $K_{n,n-1}$ of the stacking energy $W$ that have been obtained from
Ref.~\cite{Alexandrov2009ePBD}, all other parameter values we use are from Ref.~\cite{Campa1998} :
 $m=300$ amu for the base pair reduced mass, $D_{GC} = 0.075$ eV, $a_{GC} = 6.9$ \AA $^{-1}$ and
$D_{AT} = 0.05$ eV, $a_{AT} = 4.2$ \AA$^{-1}$ for G-C and A-T base pairs respectively
in the Morse potential, and $\rho = 2$, $b = 0.35$ \AA$^{-1}$. These values have been fitted to successfully reproduce specific
melting curves in short oligonucleotides. They have been extensively used in a number of prior works (for example in Refs.~\cite{Choi2004,Kalosakas2004,Voulgarakis2004,Ares2005,Alexandrov2006,Kalosakas2006,nvoul,RaptiEPL,Ares2007,Choi2008,Alexandrov2009,Kalosakas2009,Apostolaki2011,Traverso2015,Hillebrand2019,Alexandrov2010,Nowak13}). The parameters $K_{n,n-1}$
of Eq.~(\ref{eq:PBDStack}) are taking on sequence specific values (see~\cite{Alexandrov2009ePBD,Hillebrand2020})
and they have been shown to accurately reproduce peculiar denaturation transition temperatures exhibited by
homogeneous and periodic DNA oligonucleotides \cite{Alexandrov2009ePBD}.

The DNA promoters considered here are presented below. For clarity, only one strand of each sequence is shown, while the complementary strand is implied. The TSS is explicitly indicated as it is preceded by (+1).
\begin{itemize}
     \item A 69 base pair segment of the viral AAV P5 promoter: \\
     5'-GTGGCC ATTTAGGGTA TATATGGCCG AGTGAGCGAG CAGGATCTCC  (+1)ATTTTGACCG CGAAATTTGA ACG-3'
     \item A 129 base pair segment of the bacterial Lac operon promoter: \\
     5'-GAAAGCGGG CAGTGAGCGC  AACGCAATTA ATGTGAGTTA GCTCACTCAT TAGGCACCCC AGGCTTTACA CTTTATGCTT CCGGCTCGTA TGTTGTGTGG  (+1)AATTGTGAGC GGATAACAAT TTCACACAGG-3'
     \item A 81 base pair segment of the artificial superpromoter SCP1: \\
     5'-GTACTT ATATAAGGGG GTGGGGGCGC GTTCGTCCTC (+1)AGTCGCGATC GAACACTCGA  GCCGAGCAGA CGTGCCTACG GACCG-3'
\end{itemize}
Additionally to these promoters, one mutant is also examined for each case:
The mutated AAV P5 promoter is obtained by changing the base pairs at sites $+1$ and $+2$ from A-T and T-A
to G-C and C-G, respectively \cite{Choi2004,Kalosakas2004}.
In the Lac UV5 mutant the sites $-9$ and $-8$ of Lac operon
are changed from G-C and T-A base pairs both to A-T \cite{Gilbert1976}. 
Finally, in the mutant m1SCP1 of the superpromoter SCP1 the base pairs at sites $-5$ and $-4$ change from T-A and C-G
to C-G and G-C, respectively, and the sites $+8$ and $+15$ change both from A-T to G-C \cite{Alexandrov2009}. 

We have made extensive constant-energy molecular dynamics simulations using the Hamiltonian of Eq.~(\ref{eq:PBDHamiltonian})
and periodic boundary conditions.
Random initial conditions were implemented, having a fixed energy corresponding to a temperature of 310 K,
and the equations of motion were evolved using a symplectic integrator, namely the symplectic Runge-Kutta-Nystr\o m fourth order integration scheme SRKNb6 \cite{Blanes2002}. 
The threshold values of $y_{AT}^{thr}=0.24$ \AA\ and $y_{GC}^{thr}=0.15$ \AA\ are used in order to consider openings
of A-T and G-C base pairs respectively, which are derived through the characteristic lengths of the corresponding Morse potential
and are also consistent with the requirement that 50\% of the DNA chain is open at the melting transition~\cite{Hillebrand2020}.
The system is evolved for 10 ns to provide thorough thermalisation, and then base pair displacement data are recorded every picosecond for the next nanosecond.
We note that the timescales here are for our coarse-grained model.

From this displacement data, the bubble probabilities and lifetime distributions can be calculated (see Ref.~\cite{Hillebrand2020}
for the details of the procedure). Then average bubble lifetimes are computed from the corresponding lifetime distributions.
In addition to examining bubbles with a fixed length of $l=q$ base pairs (as in Ref.~\cite{Hillebrand2020}), we also take here
a more flexible approach of allowing the size of the bubble to fluctuate by considering bubbles of length $l>q$ for some value of $q$, starting at a given site. In order to reduce potential issues arising from statistical inadequacy due to the rarity of bubble occurrence,
for the calculation of bubble lifetimes we have used 10,000 simulations with different random initial conditions for each of the considered promoter sequences.

\section{Average Bubble Lifetimes in the Wild Type and Mutant Promoters} 
\label{sec:promoter_lifetimes}

We are interested here in the profiles of the average lifetimes $\langle t \rangle$ for bubbles with either a fixed length $l=q$
or length $l>q$, along the promoter sequences, as well as in the changes of these profiles with the respective mutations
that affect the transcriptional activity.
To quantify the effect of the mutations, which some of them reduce transcription while others enhance transcription,
we consider the sequence dependence of the relative difference in average bubble lifetimes between the mutated and
the wild type promoter, calculated as
\begin{equation}
    \Delta\langle t \rangle_{\textit{rel}} = \frac{\langle t \rangle_\textit{Mut} - \langle t \rangle_\textit{WT}}{\langle t \rangle_\textit{WT}}.
    \label{eq:reldiff}
\end{equation}
where $\langle t \rangle_\textit{Mut}$ corresponds to the average lifetime in the mutated promoter and $\langle t \rangle_\textit{WT}$
in the wild type. Using the relative difference $\Delta\langle t \rangle_{\textit{rel}}$ we can clearly identify regions along the sequence where the mutations increase or decrease the overall bubble lifetimes, as these areas will exhibit positive or negative relative differences respectively.

\begin{figure}[tb]
    \centering
    \includegraphics[width=0.49\textwidth]{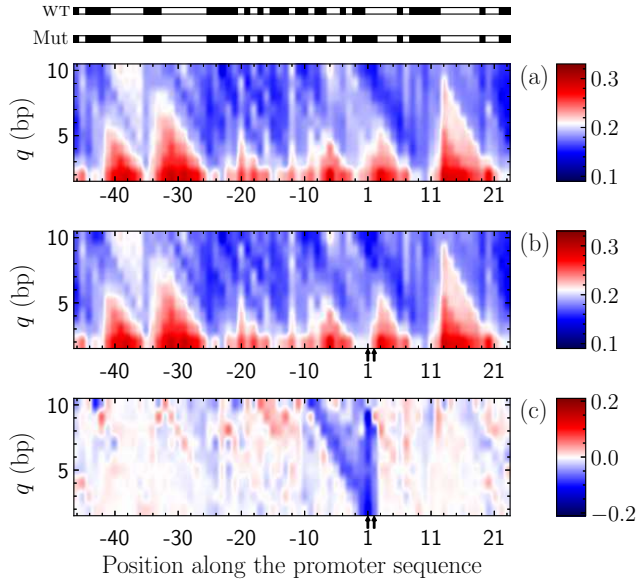}
    \caption{Average bubble lifetimes for bubbles of length $l=q$ base pairs (bp) starting at a given base pair, as a function of the position
    of this base pair along the sequence for {\bf (a)} the AAV P5 promoter and {\bf (b)} the mutated P5 promoter.
    Color bars in (a) and (b) indicate bubble lifetimes in ps.
    The horizontal bars at the top show the distribution of A-T or T-A (white) and G-C or C-G (black)
    base pairs along the wild type (WT) and the mutated (Mut) sequence, respectively. {\bf (c)} Relative difference
    $\Delta\langle t \rangle_\textit{rel}$, Eq.~\eqref{eq:reldiff}, between lifetimes of bubbles with length  $l=q$
    in the wild type and the mutated P5 promoter, as shown by the color bar at the right.
    The arrows below the x-axes in (b) and (c) indicate the mutation sites (see text).}
    \label{fig:p5_lifetimes}
\end{figure}

\begin{figure}[tb]
    \centering
    \includegraphics[width=0.49\textwidth]{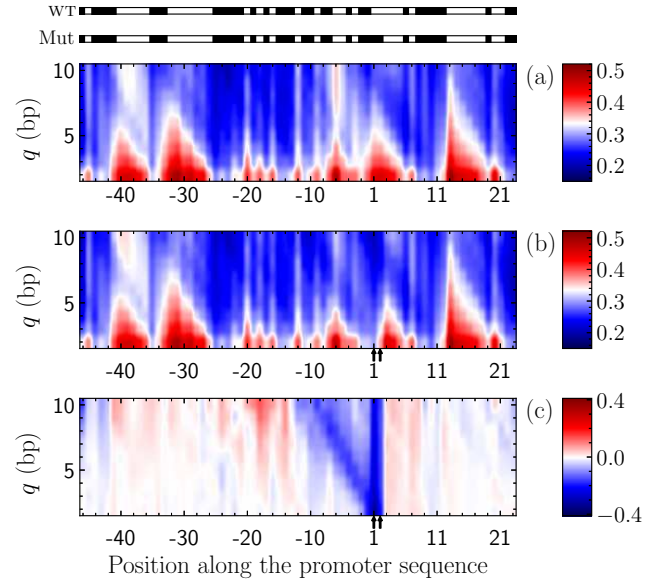}
    \caption{Average bubble lifetimes for bubbles of length $l>q$ starting at a given base pair, as a function of the position
    of this base pair along the sequence for {\bf (a)} the AAV P5 promoter and {\bf (b)} the mutated P5 promoter.
    Color bars in (a) and (b) indicate bubble lifetimes in ps.
    {\bf (c)} Relative difference $\Delta\langle t \rangle_\textit{rel}$, Eq.~\eqref{eq:reldiff}, between lifetimes of bubbles
    with length $l>q$ in the wild type and the mutated P5 promoter, as shown by the color bar at the right.
   The horizontal bars at the top, as well as the arrows below the axes in (b) and (c) are as in Fig.~\ref{fig:p5_lifetimes}.}
    \label{fig:p5_flucts}
\end{figure}

Figure~\ref{fig:p5_lifetimes} shows the profiles of the average bubble lifetimes in the AAV P5 promoter [Fig.~\ref{fig:p5_lifetimes}(a)] and
the mutated P5 promoter [Fig.~\ref{fig:p5_lifetimes}(b)], as well as the relative difference $\Delta \langle t \rangle_\textit{rel}$
between the two variants [Fig.~\ref{fig:p5_lifetimes}(c)], for different bubble lengths $l=q$, with $q$ ranging from 1 up to 10 base pairs.
We see that there are regions of more persistent bubbles both upstream and downstream of the TSS, as well as at the TSS
[Fig.~\ref{fig:p5_lifetimes}(a)], correlated with the location of A/T rich bands along the sequence (see the horizontal bars above the
plots). The effect of the TSS mutation, changing two A-T and T-A base pairs to G-C and C-G respectively
that leads to loss of transcriptional activity, shows a clear reduction of the bubble lifetime in this region [Fig.~\ref{fig:p5_lifetimes}(b)],
which is emphasised in the plot of the relative difference $\Delta \langle t\rangle_\textit{rel}$ in Fig.~\ref{fig:p5_lifetimes}(c).
Note that the color scale in Fig.~\ref{fig:p5_lifetimes}(c), and all the relative difference plots shown in the other figures  below,
is set symmetrically so that white regions correspond to zero relative difference, red regions to a positive relative difference
(i.e.~longer lifetimes in the mutated promoter) and blue regions to a negative relative difference.
Here we see a roughly 10\% decrease in bubble lifetimes for bubbles starting or ending near the mutation sites (which coincide
to the TSS), significantly reducing the lifetime of bubbles with lengths up to 10 base pairs in this region.

We have also considered bubbles of length $l>q$ for different values of $q$ up to 10 base pairs, which allow for fluctuations
of the length of the bubble without effectively destroying and recreating new bubbles constantly in the numerical simulations.
These results for the P5 promoter and the considered mutation are shown in Fig.~\ref{fig:p5_flucts}, in the same way as the results
of fixed length bubbles presented in Fig.~\ref{fig:p5_lifetimes}. It is immediately apparent that the overall lifetimes are longer
when the strictness of the fixed bubble length criterion is relaxed, with the longest average bubble lifetimes around $0.5$ ps
[Fig.~\ref{fig:p5_flucts}(a) and (b)] for fluctuation-allowed bubbles as compared to $0.3$ ps in the fixed-length case [Fig.~\ref{fig:p5_lifetimes}(a) and (b)].
The statistics of the bubble lifetime profile is better in this case since all bubbles longer than length $q$ now contribute
to the data at each length.
The same trends as in Fig.~\ref{fig:p5_lifetimes} are also apparent in Fig.~\ref{fig:p5_flucts}. The effect of the mutation is to significantly
shorten the lifetimes of bubbles near and immediately upstream of the TSS, without affecting the lifetimes along the rest of the sequence.
However,  Fig.~\ref{fig:p5_flucts}(c) shows an even stronger reduction in the average lifetimes of bubbles at the TSS due to the mutation,
at levels larger than 20\% reaching up to 30\% as $q$ decreases from 10 to 2 base pairs, suggesting a correlation between 
significant changes in bubble lifetimes and altered transcriptional activity.

As the base pair opening thresholds $y_{AT/GC}^{thr}$ considered here are relatively small as compared to previous works,
the multi-peaked bubble's inherent lifetime profiles shown in Figs.~\ref{fig:p5_lifetimes} and \ref{fig:p5_flucts}
for the AAV P5 promoter and its mutant are more reminiscent of the equilibrium average bubble probabilities~\cite{RaptiEPL},
than the out-of-equilibrium probability profiles of much larger amplitude bubbles  \cite{Kalosakas2004,Alexandrov2006},
or even the bubble lifetime distributions obtained through  Langevin dynamics~\cite{Alexandrov2006} .
However, all these works demonstrate the significant reduction of the bubble probability or relative lifetime
at the position of the mutation in the TSS.

\begin{figure}[tb]
    \centering
    \includegraphics[width=0.49\textwidth]{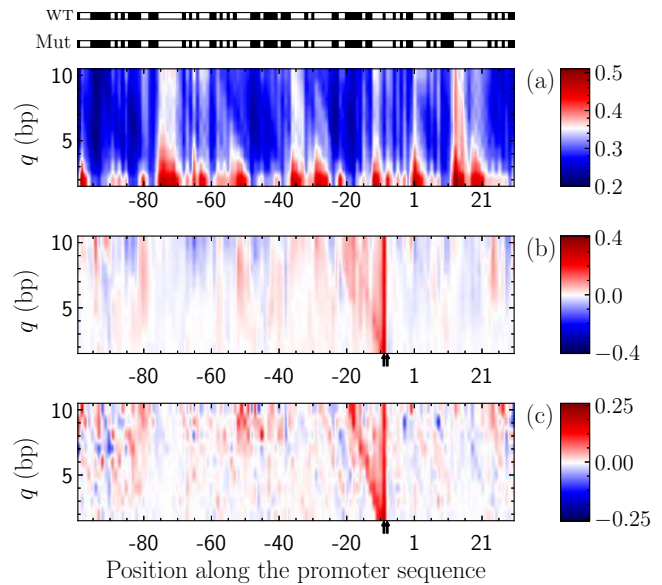}
    \caption{ {\bf (a)} Average bubble lifetime profile in the wild type Lac operon promoter for bubbles of length $l>q$ starting at a given
    base pair, as a function of the position of this base pair along the sequence. The color bar indicates bubble lifetimes in ps.
    The relative difference $\Delta\langle t \rangle_\textit{rel}$, Eq.~\eqref{eq:reldiff}, profile {\bf (b)} for bubbles of length $l>q$ and 
    {\bf (c)} for bubbles of length $l=q$. The horizontal bars at the top of the figure depict the distribution of A-T/T-A (white) and G-C/C-G (black) base pairs in the sequence for the wild type Lac operon (WT) and the Lac UV5 mutation (Mut).
    The arrows below the axes in (b) and (c) indicate the mutation sites.
    The color scale in (b) and (c) is set symmetrically, so that white regions signify no relative difference.}
    \label{fig:lac_lifetimes}
\end{figure}

Turning to the bacterial Lac operon promoter, we present the lifetimes for bubbles of length $l>q$, as well as the relative difference  between the wild type and mutant Lac UV5 in Fig.~\ref{fig:lac_lifetimes}.
The bands of longer-lived bubbles are once again correlated with A/T-dense regions of the sequence [Fig.~\ref{fig:lac_lifetimes}(a)], including the region around the TSS.
While there is no prior study on bubble lifetimes in the Lac operon, as in the previous case the obtained lifetime profile is in accordance
with equilibrium base pair opening probabilities in this promoter calculated through Monte Carlo simulations.
In particular, the main peaks of the equilibrium bubble opening propensity have been observed~\cite{Apostolaki2011} in
({\it i}) a large upstream region extending from around $-80$ up to $-50$,
({\it ii}) near the two binding sites of the polymerase subunit $\sigma$ factor at around $-30$ (a larger peak)
and in the region from $-10$ up to TSS (a smaller one), and finally ({\it iii}) downstream in the region from +10 up to +20.

In contrast to the AAV P5 mutation examined above which inhibits transcriptional activity, the Lac UV5 mutation is known to
strengthen the promoter \cite{Gilbert1976,Reznikoff1976}.
Figure~\ref{fig:lac_lifetimes}(b) shows the relative difference $\Delta\langle t\rangle_\textit{rel}$ for bubbles of length $l>q$,
while Fig.~\ref{fig:lac_lifetimes}(c) gives the corresponding profile for fixed bubble length $l=q$.
We see that at the mutations site, which is located within the -10 element of the promoter (the one of the two binding sites
of the $\sigma$ factor), Lac UV5 exhibits bubbles that tend to last longer, from 20\% up to more than 30\% in the case of $l>q$
depending on the bubble length. As previously, the relative difference is smaller, up to 25\%, for the fixed length $l=q$ bubbles.
Similarly to the bubble lifetime enhancement, there is an increase of the equilibrium opening probability at this region
in Lac UV5 as compared to the wild type profile~\cite{Apostolaki2011}.
Therefore we see that the transcription-strengthening mutation results in longer-living bubbles at the transcriptionally
functional binding site of $\sigma$ factor.

\begin{figure}[tb]
    \centering
    \includegraphics[width=0.49\textwidth]{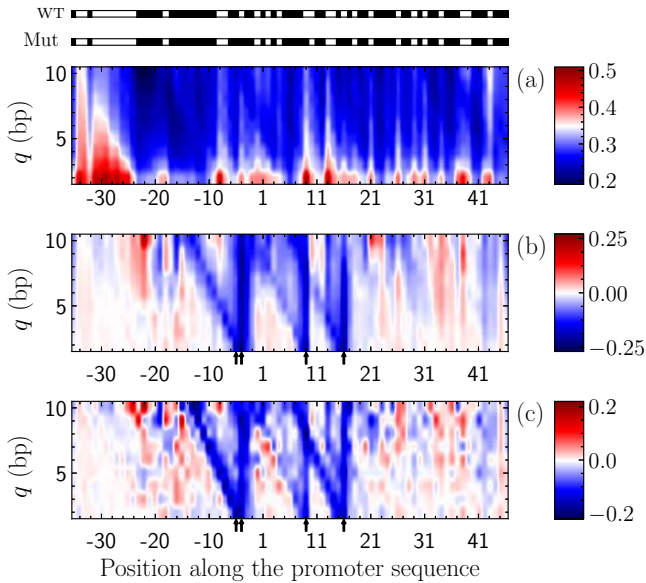}
    \caption{ {\bf (a)} Average bubble lifetimes in the SCP1 superpromoter for bubbles of length $l>q$ starting at a given
    base pair, as a function of the position of this base pair along the sequence. The color bar indicates bubble lifetimes in ps.
    The relative difference $\Delta\langle t \rangle_\textit{rel}$~\eqref{eq:reldiff} profile {\bf (b)} for bubbles of length $l>q$ and 
    {\bf (c)} for bubbles of length $l=q$.
    The horizontal bars at the top of the figure depict the distribution of A-T or T-A (white) and G-C or C-G (black) base pairs in the sequence
    for the wild type (WT) and mutation (Mut). The arrows below the axes in (b) and (c) indicate the mutation sites.
    The color scale in (b) and (c) is set symmetrically, so that white regions signify no relative difference.}
    \label{fig:scp1_lifetimes}
\end{figure}

Finally, the profile of the average bubble lifetimes for the SCP1 superpromoter is shown in Fig.~\ref{fig:scp1_lifetimes}(a) 
for bubbles with length $l>q$, while the relative differences of average bubble lifetimes 
between this promoter and its mutation m1SCP1 \cite{Alexandrov2010} are depicted in Fig.~\ref{fig:scp1_lifetimes}(b)
for the case of bubbles with variable length $l>q$ and Fig.~\ref{fig:scp1_lifetimes}(c) for bubbles with fixed length $l=q$.
The equilibrium probabilities for large amplitude bubbles show a large peak at the region further upstream from the
position $-30$ and another smaller peak around the TSS \cite{Alexandrov2010}. In agreement with these observations
are the dominant feature around $-30$ in our average lifetime profiles as well as the peak around TSS [Fig.~\ref{fig:scp1_lifetimes}(a)].
However, regarding small amplitude bubbles, as those considered in our case, there are additional peaks further downstream
of the TSS, that are not present for bubbles with amplitude larger than 3.5 \AA\ in equilibrium \cite{Alexandrov2010}.

A primary finding obtained by Langevin dynamics simulations of SCP1 superpromoter was a region immediately
downstream from the TSS (located between +1 and +10) where large long-lived bubbles tended to form, while the introduction
of the transcription-inhibiting mutations of m1SCP1 led to the destruction of this dominant peak  \cite{Alexandrov2010}.
Our results also show a substantial decrease in the average bubble lifetimes at this region downstream from the TSS,
that becomes more clear for larger bubble lengths especially in the $l>q$ case [Fig.~\ref{fig:scp1_lifetimes}(b)].
Once more, the relative differences of the average bubble lifetimes around the mutated sites show a decrease, more than 20\%
for smaller lengths when bubbles of fluctuating ends are considered ($l>q$ ), for the m1SCP1 mutant that suppress
transcription as compared to the SCP1 superpromoter.

\section{Conclusions} 
\label{sec:conclusions}

Using the Peyrard-Bishop-Dauxois coarse-grained model with sequence dependent stacking interactions for the description of base pair openings in DNA
and efficient numerical techniques, we have performed extended microcanonical simulations
to investigate average bubble lifetime profiles along the sequence of  three different promoters,
the viral AAV P5, the bacterial Lac operon, and the artificial SCP1 superpromoter, as well as one mutation for each promoter.
The time scales of the inherent DNA double strand transient separations have been probed in the framework of this model,
with no artificial time scales imposed through arbitrary friction coefficient, and using a physically motivated base-pair-dependent
threshold value for considering base pairs to be open.
The inherent bubble lifetimes, for relatively small amplitude bubbles of the order of tenths of \AA, in our constant energy
simulations are in the subpicosecond time scale, as opposed to Langevin fluctuational dynamics computations revealing
bubble lifetimes of the order of picosecond for larger amplitude bubbles of a few \AA.

We found that transcription-inhibiting mutations in the case of the AAV P5 and SCP1 promoters resulted in significant reductions of bubble lifetimes around the transcriptionally relevant mutated sites, while transcription-boosting mutations of the Lac operon promoter at a transcription factor binding site showed significant enhancement of the bubble lifetimes.
The corresponding negative or positive relative differences of the average bubble lifetimes between the mutated and the
wild type promoter at the position of mutations range from 20\% for larger bubble lengths up to more than 30\% for shorter ones.

\section*{Acknowledgments} 

M.~H.~acknowledges support by the National Research Foundation (NRF) of South Africa (Grant Numbers: 129630). G.~K.~and Ch.~S.~were supported by the Erasmus+/International Credit Mobility KA107 program. We thank the Center for High Performance Computing of South Africa for providing computational resources for this project.

\nocite{*}


\end{document}